\documentclass[journal,comsoc] {IEEEtran}
\usepackage{amsmath,amsfonts,amssymb}
\usepackage{graphicx}
\usepackage{xcolor}
\usepackage{subfig}
\usepackage{makecell}
\usepackage{multirow}
\usepackage{tikz}
\usepackage{mathrsfs}
\usetikzlibrary{calc,arrows,shapes,chains,positioning,shapes.geometric}
\usepackage{enumitem}
\usepackage{algpseudocode}
\usepackage{algorithm}

\usepackage{amsthm,bm}
\usepackage{footnote}
\usepackage{pgfplots}
\pgfplotsset{width=10cm,compat=1.9}
\pgfplotsset{every axis legend/.style={
    cells={anchor=center},
    inner xsep=3pt,inner ysep=2pt,
    nodes={inner sep=2pt,text depth=0.1em},
    anchor=north east,
    shape=rectangle,
    fill=white,draw=green,font=\footnotesize,
}}
\usepackage{booktabs}
\usepackage{fancyhdr}
\usepackage{svg}

\fancypagestyle{plain}{
    \fancyhf{}
    \fancyfoot[C]{\thepage}
    
}

\usepackage{hyperref}
\hypersetup{
    anchorcolor=black, 
    bookmarksnumbered=true,
    bookmarksopen=true,
    bookmarksopenlevel=1,
    colorlinks=true,
    linkcolor=blue, 
    citecolor=red,
    urlcolor=magenta
}
\begin{document}
\thispagestyle{fancy}
\title{Effective Application of Normalized Min-Sum Decoding for Short BCH Codes}
\author{Guangwen Li, Xiao Yu
\thanks{G.Li is with  Shandong Technology and Business University, Yantai, China}
\thanks{X.Yu is with  Binzhou Medical University, Yantai, China.}
}

\maketitle
\begin{abstract}
This paper introduces an enhanced normalized min-sum decoder designed to address the performance and complexity challenges associated with developing parallelizable decoders for short BCH codes in high-throughput applications. The decoder optimizes the standard parity-check matrix using heuristic binary summation and random cyclic row shifts, resulting in a Tanner graph with low density, controlled redundancy, and minimized length-4 cycles. The impact of row redundancy and rank deficiency in the dual code's minimum-weight codewords on decoding performance is analyzed. To improve convergence, three random automorphisms are applied simultaneously to the inputs, with the resulting messages merged at the end of each iteration. Extensive simulations demonstrate that, for BCH codes with block lengths of 63 and 127, the enhanced normalized min-sum decoder achieves a 1–2 dB performance gain and 100× faster convergence compared to existing parallel and iterative decoders. Additionally, a hybrid decoding scheme is proposed, which selectively activates order statistics decoding when the enhanced normalized min-sum decoder fails. This hybrid approach is shown to approach maximum-likelihood performance while retaining the advantages of the normalized min-sum decoder across a broad SNR range.
\end{abstract}


\thispagestyle{fancy}
\section{Introduction}
\label{intro_sec}
\IEEEPARstart{F}{or} low-density parity-check (LDPC) codes \cite{gallager62}, belief propagation (BP) \cite{mackay96} and its variants dominate decoding due to their near-maximum-likelihood (ML) performance and high data throughput enabled by parallelizable implementations. However, applying BP effectively to classical linear block codes with high-density parity-check (HDPC) matrices, such as Bose–Chaudhuri–Hocquenghem (BCH) and Reed–Muller (RM) codes, remains challenging. The abundance of short cycles in their Tanner graphs (TGs) disrupts the message independence required for optimal decoding, spurring significant interest in developing efficient decoding schemes for HDPC codes - particularly for ultra-reliable low-latency communication (URLLC) scenarios where both reliability and throughput are critical.

The importance of exploiting code structure to facilitate decoding has long been recognized. Jiang et al. \cite{jiang2004iterative} proposed stochastic shifting mechanisms and adaptive damping coefficients for log-likelihood ratio (LLR) updates. They further modified the parity-check matrix (PCM) $\mathbf{H}$ based on reliability updates of codeword bits across iterations \cite{jiang2006iterative}, aiming to prevent BP from stalling at pseudo-equilibrium points. Halford et al. \cite{halford2006random} introduced random redundant decoding (RRD), demonstrating improved bit error rate (BER) performance with a cycle-reduced $\mathbf{H}$. Ismail et al. \cite{ismail2015efficient} proposed permuted BP (PBP), which applies random automorphisms to messages in each iteration. Santi et al. \cite{santi2018decoding} incorporated minimum-weight parity checks tailored to received sequences, achieving stronger BP decoding for RM codes at the cost of batch processing capability. Babalola et al. \cite{babalola2019generalized} introduced a generalized parity-check transformation (GPT) to update bit reliability using on-the-fly syndrome, though this method suffers from poor BER and lacks parallelizability.

The modified RRD (mRRD) algorithm \cite{dimnik2009improved} fixed the damping factor and employed multiple parallel decoders based on $\mathbf{H}$ permutations, offering lower complexity than \cite{hehn2010multiple}, which used multiple $\mathbf{H}$s composed of cyclic shifts of minimum-weight dual-code codewords. Geiselhart et al. \cite{geiselhart2021automorphism} generalized ensemble decoding by incorporating diverse constituent decoders, achieving near-ML performance for RM codes.

Inspired by deep learning, Nachmani et al. \cite{nachmani18} proposed neural BP (NBP), which weights messages passed in TGs. Lian et al. \cite{lian2019learned} demonstrated that shared parameters tailored to varying signal-to-noise ratios (SNRs) in NBP reduce complexity without performance loss. Buchberger et al. \cite{buchberger2020pruning} pruned uninformative check nodes in overcomplete $\mathbf{H}$s for NBP. However, the performance improvement of these NBP variants in terms of frame error rate (FER) or BER remains modest relative to their complexity.

Recognizing the critical impact of $\mathbf{H}$ on BP performance, Lucas et al. \cite{lucas1998iterative} advocated $\mathbf{H}$ composed of minimum-weight dual-code codewords for their low density of non-zero elements. Yedidia et al. \cite{yedidia2002generating} proposed expanding $\mathbf{H}$ with auxiliary "bits" to reduce row weight and short cycles. Kou et al. \cite{kou01} demonstrated that redundant $\mathbf{H}$s can greatly enhance BP decoding of finite LDPC algebraic-geometry codes, inspiring similar strategies for HDPC codes.

On the other hand, universal ordered statistics decoding (OSD) \cite{Fossorier1995} variants can decode without leveraging specific code properties. Bossert et al. \cite{bossert2022hard} employed formation set decoding as an OSD variant for BCH codes, yielding competitive FER  at the cost of throughput due to OSD's serial processing nature. Similarly, the hybrid of hard-decision algebraic decoding and OSD proposed by Bailon et al. \cite{bailon2022concatenated} suffers from frequent OSD post-processing.

As such, a new decoder is urged to address challenges such as the pronounced gap to ML performance, high complexity of ensemble decoding \cite{dimnik2009improved,hehn2010multiple,geiselhart2021automorphism}, nested loops in BP decoding \cite{jiang2006iterative,ismail2015efficient,dimnik2009improved}, and limited throughput due to Gaussian elimination on $\mathbf{H}$ \cite{santi2018decoding,babalola2019generalized}. The main contributions include:
\begin{itemize}
    \item[*] A heuristic method to systematically construct an optimized $\mathbf{H}_o$ from the standard $\mathbf{H}$, achieving lower density, modest redundancy, cycle reduction, and quasi-regularity.
    \item[*] An enhanced normalized min-sum (NMS) decoder that applies multiple automorphisms to received sequences, enabling fast convergence through message aggregation.
    \item[*] A hybrid decoder combining NMS and OSD components, where NMS handles the majority of decodings due to its superior FER over parallelizable alternatives while maintaining high throughput, with OSD selectively applied  to preserve both reliability and latency requirements.
\end{itemize}

The remainder of this paper is organized as follows: Section \ref{preliminary} reviews NMS, OSD and mRRD variants. Section \ref{motivations} details the optimization of $\mathbf{H}$ and the design of an NMS decoder tailored to BCH codes. Section \ref{simulations} presents simulation results and analysis. Concluding remarks are provided in Section \ref{conclusion}.
\section{Preliminaries}
\label{preliminary}

Consider a binary message row vector $\mathbf{m} = [m_i]_1^K$, encoded into a codeword $\mathbf{c} = [c_i]_1^N$ using $\mathbf{c} = \mathbf{mG}$ over the Galois field GF(2), where $K$ and $N$ denote the lengths of the message and codeword, respectively, and $\mathbf{G}$ is the generator matrix. Each bit $c_i$ is mapped to an antipodal symbol via $s_i = 1 - 2c_i$ using binary phase-shift keying (BPSK) modulation and transmitted over a noisy channel. The decoder receives the sequence $\mathbf{y} = [y_i]_1^N$, where $y_i = s_i + n_i$, and $n_i$ is additive white Gaussian noise (AWGN) with zero mean and variance $\sigma^2$.

The $i$-th element of the log-likelihood ratio $L(\mathbf{y})$ is calculated as:
\begin{equation}
\label{reliability_def}
{L_{v_i}} = \log \left( \frac{{p(y_i|{c_i} = 0)}}{{p(y_i|{c_i} = 1)}} \right) = \frac{{2y_i}}{{\sigma^2}}.
\end{equation}
Thus, a larger magnitude of $y_i$ implies greater confidence in the hard decision of its corresponding bit, a property leveraged by OSD decoding. Unlike standard BP decoding, the NMS decoder allows arbitrary evaluation of $\sigma^2$ (e.g., $\sigma^2 = 2$) without impacting decoding performance.

\subsection{Original NMS Decoder}
The Tanner graph (TG) of a code is determined by its  $\mathbf{H}$ of size $M \times N$, where $M \geq N-K$. A variable node $v_i$ ($i = 1, 2, \ldots, N$) exchanges messages with a check node $c_j$ ($j = 1, 2, \ldots, M$) if the entry in the $j$-th row and $i$-th column of $\mathbf{H}$ is nonzero.

Under a flooding schedule in NMS decoding, the message from $v_i$ to $c_j$ at the $t$-th iteration ($t = 1, 2, \ldots, I_m$) is:
\begin{equation}
x_{v_i \to c_j}^{(t)}  = {L_{v_i}} + \sum\limits_{\substack{k \in \mathcal{C}(i)\backslash j}} {x_{c_k \to v_i}^{(t - 1)}},
\label{eq_v2c}
\end{equation}
while the message from $c_j$ to $v_i$ is updated by:
\begin{equation}
x_{c_j \to v_i}^{(t)} = \alpha \cdot \prod\limits_{\substack{k \in \mathcal{V}(j)\backslash i}} \text{sgn}\left( x_{v_k \to c_j}^{(t)} \right) \cdot \min_{\substack{ k \in \mathcal{V}(j)\backslash i}} \left| x_{v_k \to c_j}^{(t)} \right|,
\label{eq_c2v}
\end{equation}
where $\alpha$ is a normalization factor optimized via neural networks, $\mathcal{C}(i)\backslash j$ denotes the indices of neighboring check nodes of $v_i$ excluding $j$, and $\mathcal{V}(j)\backslash i$ denotes the indices of neighboring variable nodes of $c_j$ excluding $i$. All terms $x_{c_k \to v_i}^{(0)}$ in \eqref{eq_v2c} are initialized to zero. Messages alternate along the edges of the TG via \eqref{eq_v2c} and \eqref{eq_c2v} until the maximum number of iterations $I_m$ is reached. Meanwhile, the LLR output of the $i$-th bit at the $t$-th iteration is:
\begin{equation}
x_{i}^{(t)} = {L_{v_i}} + \sum\limits_{\substack{k \in \mathcal{C}(i)}} {x_{{c_k} \to {v_i}}^{(t - 1)}},
\label{eq_bit_decision}
\end{equation}
on which a tentative hard decision $\hat{\mathbf{c}} = [\hat{c}_i^{(t)}]_1^N$ is obtained and checked against the early stopping criterion:
\begin{equation}
\label{termination_criterion}
\mathbf{H}\hat{\mathbf{c}}^{\top}=\mathbf{0}.
\end{equation}

\subsection{OSD and Its Variants}
An order-$p$ OSD typically consists of the following steps:
\begin{itemize}
    \item Reliability ordering: All bits of $\mathbf{y}$ are reordered in ascending order of reliability, measured by the magnitude of \eqref{reliability_def} or an improved version \cite{bossert2022hard}.
    \item Gaussian elimination: The original $\mathbf{H}$ (assuming full rank with no redundancy) is reordered according to the same reliability ordering. Gaussian elimination is then performed to transform the reordered $\mathbf{H}$ into systematic form, which may permute the associated bits. The most reliable basis (MRB) corresponds to the last $K$ bit positions.
    \item List decoding: A list of candidate codewords is generated by flipping at most $p$ bits in the MRB, and the codeword with the highest likelihood to $\mathbf{y}$ is selected as the optimal.
\end{itemize}
As a soft-decision decoder, conventional OSD can achieve near-ML performance for high orders of $p$, but at the cost of exponentially increased complexity. Among its variants, \cite{bailon2022concatenated} proposed narrowing the search space by assigning different orders to subsections of the MRB to reduce complexity. Given its simplicity and the sufficiency of low-order OSD (e.g., $p = 2$) for the tested codes, we adopt conventional OSD as the second component of the hybrid decoder, even though variants like \cite{bossert2022hard,bailon2022concatenated} may offer better alternatives.

\subsection{mRRD and PBP Decoders}
The automorphism group of a code refers to all coordinate permutations that preserve the code. Thus, any automorphism $\rho$ belonging to BCH codes can be utilized to aid decoding. The mRRD scheme \cite{dimnik2009improved} deploys $I_3$ decoders in parallel to achieve diversity gain. For each decoder, random automorphisms are applied every $I_1$ BP iterations for a total of $I_2$ rounds. The final decision is made among candidate codewords that satisfy \eqref{termination_criterion}. In contrast, the PBP \cite{ismail2015efficient} abandons the ensemble configuration of \cite{dimnik2009improved} and applies one automorphism per BP iteration, terminating decoding immediately when \eqref{termination_criterion} is met to reduce complexity and latency.
\section{Motivations and Initiatives}
\label{motivations}

\begin{figure}[htbp]
    \centering
    \includegraphics[width=0.9\linewidth]{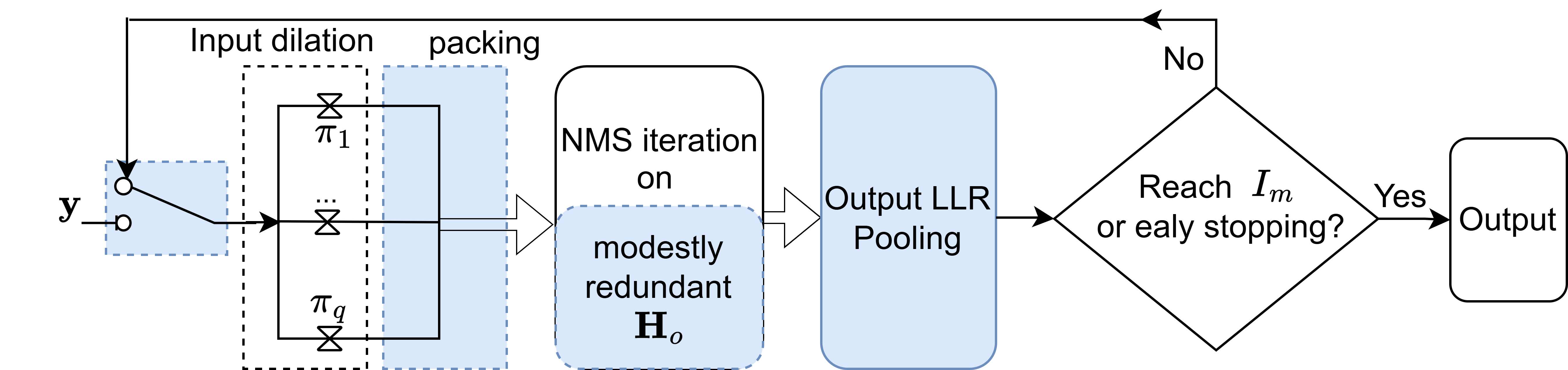}
    \caption{Block diagram of the enhanced NMS decoding.}
    \label{fig:flowchar_bch}
\end{figure}

As depicted in Fig.~\ref{fig:flowchar_bch}, the colored blocks highlight the optimized $\mathbf{H}_o$ and the specific revisions to the conventional NMS decoding.

\subsection{Choice of Parity-Check Matrix $\mathbf{H}_o$}
Regarding the adaptation of the standard $\mathbf{H}$, prior studies primarily focused on reducing its density \cite{baldi2008iterative, helmling19}, minimizing length-4 cycles \cite{halford2006random}, or increasing row redundancy \cite{baldi2008iterative}. However, a systematic scheme balancing these factors remains absent, motivating us to propose the following process for obtaining an optimized $\mathbf{H}_o$:
\begin{enumerate}
    \item Initialize $S_f = S_g = \varnothing$, $||\cdot||$ denotes the Hamming weight.
    \item Transform $\mathbf{H}$ into its row echelon form $\mathbf{H}_{r}$.
    \item Minimize $\mathbf{H}_{r}$'s density via binary additions on rows $\mathbf{r}_i$ and $\mathbf{r}_j$ to obtain $\mathbf{H}_{r_1}$. Specifically:
    
    \textbf{for} $i=1,2,\ldots,M_r$ (number of $\mathbf{H}_r$'s rows)
    
    \hspace*{0.4cm}$w_g = ||\mathbf{r}_i||; \; S_f \gets \left\{ \mathbf{r}_i \right\}$
    
    \hspace*{0.4cm}\textbf{for} $j=1,2,\ldots,M_r, \; j\neq i$
    
    \hspace*{1.0cm}$\mathbf{t} = \mathbf{r}_i \oplus \mathbf{r}_j$
    
    \hspace*{1.0cm}\textbf{if} $||\mathbf{t}|| = w_g$, $S_f \gets S_f \cup \left\{ \mathbf{t} \right\}$
    
    \hspace*{1.0cm}\textbf{if} $||\mathbf{t}|| < w_g$, $S_f \gets \left\{ \mathbf{t} \right\}; \; w_g \gets ||\mathbf{t}||$
    
    \hspace*{0.4cm}$S_g \gets S_g \cup S_f$
    
    Construct $\mathbf{H}_{r_1}$ with the elements of $S_g$, excluding any cyclically shifted versions of existing rows.
    \item Perform an exhaustive search (e.g., $Q = 4$ times):
    
    \textbf{for} $i=1,2,\ldots,Q$
    
    \hspace*{0.3cm}$\mathbf{H}_r\gets \mathbf{H}_{r_1}$; Repeat step 2, replacing $\mathbf{r}_j$ with each of its cyclic shifts $\mathbf{r}_j^{(q)}$, $q = 1,2,\ldots,N$.
    \item For $\mathbf{H}_{r_1}$, if its row count $M_{r_1} \le M$, pad it with $\beta(M - M_{r_1})$ additional rows of $\mathbf{H}_{r_1}$ in ascending order of weight, where $\beta$ denotes the redundancy factor.
    \item Apply a heuristic method (e.g., simulated annealing) to probabilistically select a row based on the number of its associated length-4 cycles. Impose a random shift on it to minimize a loss function defined by the total number of length-4 cycles and column weight variance. This results in the optimized $\mathbf{H}_o$. $\hspace{1cm}\blacksquare$
\end{enumerate}

The BCH codes of varying rates with block lengths 63 and 127 have been validated using the above procedures. As demonstrated in Table~\ref{tab:matrix-table}, the number of length-4 cycles and the standard deviation of column weight are effectively reduced, even with modest row redundancy introduced for $\mathbf{H}_o$.

\begin{table}[!t]
\caption{\scriptsize \uppercase{Attributes of chosen parity-check matrices for BCH codes of varying rates with block lengths 63 and 127.}}
\label{tab:matrix-table}
\resizebox{0.49\textwidth}{!}
{
\begin{tabular}{|c|c|c|c|c|c|}
\hline
Code &
  \begin{tabular}[c]{@{}c@{}}PCM \\ (dimensions)\end{tabular} &
  \begin{tabular}[c]{@{}c@{}}\# of\\ length-4 cycles\end{tabular} &
  \begin{tabular}[c]{@{}c@{}}Column weight\\ Range/Mean/Std\end{tabular} &
  \begin{tabular}[c]{@{}c@{}}Row weight\\ Range/Mean/Std\end{tabular} &
  $\beta$ \\ \hline
\multirow{3}{*}{(63,36)}  & $\mathbf{H}$(27$\times63$) \cite{helmling19}     & 5909   & {[}1,13{]} / 7.7 / 3.9   & {[}18,18{]} / 18 / 0.0   & -  \\ \cline{2-6} 
                          & $\mathbf{H}_o$(32$\times$63)  & 2521   & {[}7,10{]} / 7.9 / 0.9   & {[}14,16{]} / 15.5 / 0.9 & 2  \\ \cline{2-6} 
                          & \color{blue}$\mathbf{H}_o$(122$\times$63) &\color{blue} 42724  &\color{blue} {[}27,31{]} / 29.7 / 1.2 &\color{blue} {[}14,16{]} / 15.3 / 0.9 &\color{blue}\textbf{20}  \\ \hline
\multirow{3}{*}{(63,45)}  & $\mathbf{H}$(18$\times$63) \cite{helmling19}      & 7251   & {[}1,11{]} / 6.9 / 2.9   & {[}24,24{]} / 24 / 0.0   & -  \\ \cline{2-6} 
                          & \color{blue}$\mathbf{H}_o$(33$\times$63)  & \color{blue}3066   & \color{blue}{[}7,11{]} / 8.4/ 1.0    &\color{blue} {[}16,16{]} / 16 / 0.0   & \color{blue}\textbf{2}  \\ \cline{2-6} 
                          & $\mathbf{H}_o$(78$\times$63)  & 19620  & {[}17,21{]} / 19.8 / 1.0 & {[}16,16{]} / 16 / 0.0   & 5  \\ \hline
\multirow{2}{*}{(127,64)} & $\mathbf{H}$(63$\times$127) \cite{helmling19}     & 138779 & {[}1,33{]} / 16.9 / 9.6  & {[}34,34{]} / 34 / 0.0   & -  \\ \cline{2-6} 
                          & \color{blue}$\mathbf{H}_o$(72$\times$127) &\color{blue} 14900  &\color{blue} {[}10,15{]} / 12.6 / 1.1  & \color{blue}{[}22,24{]} / 22.2 / 0.6 & \color{blue}\textbf{2}  \\ \hline
\multirow{2}{*}{(127,78)} & $\mathbf{H}$(49$\times$127) \cite{helmling19}     & 205240 & {[}1,31{]} / 17.0 / 8.8  & {[}44,44{]} / 44 / 0.0   & -  \\ \cline{2-6} 
                          & \color{blue}$\mathbf{H}_o$(57$\times$127) & \color{blue}24807  & \color{blue}{[}9,15{]} / 12.6 / 1.1  & \color{blue}{[}28,28{]} / 28 / 0.0   & \color{blue}\textbf{2} \\ \hline
\multirow{2}{*}{(127,99)} & $\mathbf{H}$(28$\times$127) \cite{helmling19}     & 66166  & {[}1,15{]} / 10.6 / 3.4  & {[}48,48{]} / 48 / 0.0   & -  \\ \cline{2-6} 
                          & \color{blue}$\mathbf{H}_o$(33$\times$127) & \color{blue}53278  & \color{blue}{[}9,15{]} / 11.4 / 1.2  & \color{blue}{[}44,44{]} / 44 / 0.0   & \color{blue}\textbf{2}  \\ \hline
\end{tabular}
}
\end{table}

\begin{figure}[htbp]
    \centering 
    \includegraphics[width=0.9\linewidth]{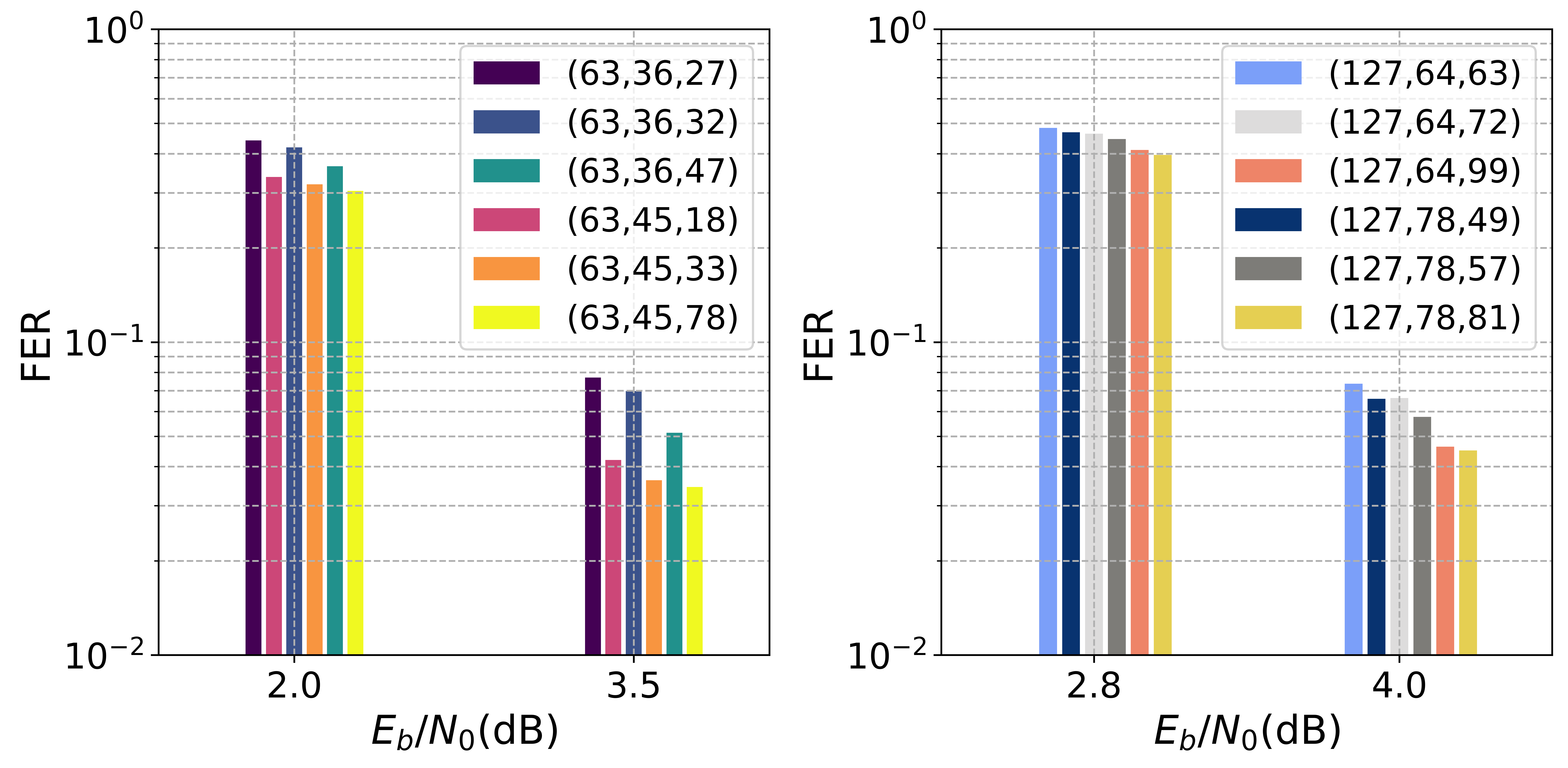}
    \caption{FER of BCH codes with varying $\beta$ in $\mathbf{H}_o$.}
    \label{fig:fer127_63codes}
\end{figure}

The row number $M_o$ of $\mathbf{H}_o$ is influenced by the redundancy factor $\beta$, which impacts the FER metric of the enhanced NMS. For the BCH (63,36) code, fixing the empirically optimal $\alpha$ for the enhanced NMS with $I_m=4$ at SNR = 2.6 dB, the ($\beta,M_o,\text{FER}$) at SNR = 3.5 dB (assuming no knowledge of noise variance) evolves as $(1,27,0.077) \to (5,47,0.051) \to (20,122,0.030) \to (50,269,0.027) \to (60,316,0.028)$. The FER improves initially with increasing $\beta$, but diminishing returns are observed until $\beta$ reaches 50, after which further increases yield worse FER. Similar trends are observed for codes of varying rates and other block lengths, as illustrated in Fig.~\ref{fig:fer127_63codes}, where legend entries are presented in the form $(N,K,M_o)$ (some $\beta > 5$ cases omitted for clarity). However, the evaluation of the optimal $\beta$ depends on factors such as the implementation of the enhanced NMS with tunable parameters ($\alpha, I_m$) and the SNR points under test. Additionally, the computational cost of high $\beta$ values necessitates a trade-off between FER and complexity.

Another intriguing observation is the comparison of FERs between codes with $\mathbf{H}_o$ of similar size. Specifically, the (63,45) code outperforms the (63,36) code despite its higher rate and slightly smaller $M_o$. A similar trend is observed for the (127,78) code compared to the (127,64) code. Further analysis reveals that the full-rank $\mathbf{H}_o$s for (63,36) and (127,64) codes are composed of rows with varying Hamming weights, whereas the (63,45) and (127,78) codes have full-rank $\mathbf{H}_o$s composed of uniform minimum-weight rows (16 and 28, respectively). The property of rank deficiency—the inability to compose a full-rank $\mathbf{H}_o$ with uniform minimum-weight rows—has been overlooked in the literature but plays a critical role in explaining why (63,45) and (127,78) codes outperform their counterparts in FER, despite having denser matrices, less redundancy, and more length-4 cycles. Notably, this discussion may apply only to BP-like decoders.

\subsection{Enhanced NMS Decoder}
Recognizing that automorphisms of the received sequence $\mathbf{y}$ can represent different interpretations of the same transmitted codeword \cite{baldi2008iterative}, we propose dilating the input by aggregating these representations into an input block $B(\mathbf{y})$. Three types of automorphisms are utilized: 
\begin{itemize}
    \item Interleaving $\pi_{_I}$, which concatenates bits at even indices with those at odd indices.
    \item Frobenius mapping $\pi_{_F}$, which maps bit index $i \mapsto 2i \mod N$.
    \item Cyclic shifting $\pi_{_C{_s}}$, defined as $s \cdot d_p + d_o \mod N$, where $d_p = 21$ and $42$ for length-63 and 127 codes, respectively, $s \in S_n = \{0,1,2\}$, and a random offset $d_o \in [0, d_p)$.
\end{itemize}
For $s \in S_n$, apply $\pi_{_C{_s}}$ to $S_p = \{\mathbf{y}, \pi_{_I}(\mathbf{y}), \pi_{_F}(\mathbf{y})\}$ individually to yield a block $B(\mathbf{y})$ of size $|S_n||S_p| = 9$.

As illustrated in Fig.~\ref{fig:flowchar_bch} and Algorithm~\ref{alg::enhanced_nms}, the enhanced NMS decoder largely retains the structure of the original NMS, with alterations in input and output processing at each iteration. Specifically, at step 2, the input is dilated by automorphisms to support decoding; at step 4, the output is aligned and averaged before applying \eqref{eq_bit_decision}; at step 5, the original $\mathbf{y}$ is replaced by $\mathbf{y}^{(t)}$ after the first iteration.

\begin{algorithm}[ht]
\caption{\hspace{1cm}Enhanced NMS Decoder}
\label{alg::enhanced_nms}
\begin{algorithmic}[1]
\Require
Received $\mathbf{y}$, 
$\mathbf{H}_o$, and three types of permutations.
\Ensure
Optimal codeword estimate $\hat{\mathbf{c}}$ for $\mathbf{y}$.
\State \textbf{for} $t = 1, 2, \ldots, I_m$
\State \hspace{0.3cm} Dilate the received sequence into a block $B(\mathbf{y})$ with shape $|S_n||S_p| \times N$, where $|\cdot|$ denotes set cardinality.
\State \hspace{0.3cm} Perform the variable-to-check message passing in \eqref{eq_v2c} for $B(\mathbf{y})$, followed by the check-to-variable update in \eqref{eq_c2v}.
\State \hspace{0.3cm} Reverse the permutations applied to the output of \eqref{eq_c2v}, average the results, and substitute them for the second term in the RHS of \eqref{eq_bit_decision} to compute $\mathbf{y}^{(t)}$ and its hard decision $\hat{\mathbf{c}}^{(t)}$.
\State \hspace{0.3cm} \textbf{if} $\mathbf{H}_o \hat{\mathbf{c}}^{(t)^\top} \ne \mathbf{0}$ \textbf{then} update $\mathbf{y} \gets \mathbf{y}^{(t)}$.
\State \hspace{0.3cm} \textbf{else} \textbf{break}.
\State Return $\hat{\mathbf{c}} = \hat{\mathbf{c}}^{(t)}$.
\end{algorithmic}
\end{algorithm}

The main characteristics of mRRD and PBP decodings include standard BP procedures, three nested loops, and one automorphism on the input per iteration. In contrast, the enhanced NMS adopts simplified NMS core operations, a single loop, and multiple automorphisms on the input per iteration. It has been found that the enhanced NMS accelerates decoding convergence, justifying the use of a small $I_m$.
\section{Simulation Results and Analysis}
\label{simulations}

Five BCH codes are evaluated: (63,36), (63,45), (127,64), (127,78), and (127,99) \cite{helmling19}. For the enhanced NMS (referred to as NMS), $I_m = 4$ for codes of block length 63 and $I_m = 8$ for block length 127. The normalization factor $\alpha$ is tuned and fixed after training the NMS as a neural network at SNR = 2.6 dB and 3.0 dB for block lengths 63 and 127, respectively. The redundancy factor $\beta$ for $\mathbf{H}_o$ is set to 20 for the (63,36) code and 2 for the other codes, as shown in Table~\ref{tab:matrix-table}. Open-source code for reproduction is available on GitHub\footnote{\url{https://github.com/lgw-frank/Short\_BCH\_Decoding\_OSD}}, implemented on the TensorFlow platform. For fairness, decoding results from the literature were calibrated directly from their respective sources rather than re-implementing the decoders.

\subsection{Decoding Performance}

\begin{figure}[!t]
    \centering
    \subfloat[BER of (63,36) code\label{fig:ber63_36}]{
    \resizebox{0.49\linewidth}{!}{ 
	\begin{tikzpicture}[scale=0.55]
		\begin{semilogyaxis}[
			scale = 0.75,
			xlabel={$E_b/N_0$(dB)},
			ylabel={BER},
			xmin=1.0, xmax=4.55,
			ymin=1e-6, ymax=0.12,
			xtick={0.0,0.5,1.0,1.5,2,...,4.0,4.5},
			legend pos = south west,
			ymajorgrids=true,
			xmajorgrids=true,
			grid style=dashed,
			legend style={legend columns=1,fill opacity=0.6},
            draw opacity=1,text opacity=1,
                xminorgrids=false, 
                yminorgrids=true,
			]
\addplot[
color=black,
mark=diamond*,
solid,
very thin
]
coordinates {
(1.5,0.09)
(2.0,0.075)
(3.0,0.04)
(4.0,0.018)
(5.0,5.1e-3)
};	
\addlegendentry{GPT(10)\cite{babalola2019generalized}}

\addplot[
color=teal,
mark=square,
very thin
]
coordinates {
(1.0,1.1e-1)
(2.0,7e-2)
(3.0,4e-2)
(4.0,1.35e-2)
(5.0,3.3e-3)
};
\addlegendentry{ BP-RNN \cite{nachmani18}} 
\addplot[
color=red,
mark=triangle,
solid,
very thin
]
coordinates {
(2.0,0.048)
(2.5,0.023)
(3.0,0.011)
(3.5,0.0042)
(4.0,1.3e-3)
(4.5,3.6e-4)
(5.0,1e-4)
};	
\addlegendentry{RNN-SS(RRD)\cite{lian2019learned}}

\addplot[
color=violet,
mark=halfcircle,
solid,
very thin
]
coordinates {
(0.0, 0.12944)
(0.5, 0.10649)
(1.0, 0.08127)
(1.5, 0.05508)
(2.0, 0.03437)
(2.2, 0.02908)
(2.4, 0.02222)
(2.6, 0.01823)
(2.8, 0.01264)
(3.0, 0.00909)
(3.2, 0.0065)
(3.4, 0.00423)
(3.6, 0.00283)
(3.8, 0.0018)
(4.0, 0.00104)
(4.2, 0.00064)
(4.4, 0.00036)
};	
\addlegendentry{NMS(4)}

\addplot[
color=orange,
mark=diamond,
very thin
]
coordinates {
(3.0,4e-3)
(3.5,1.2e-3)
(4.0,2.4e-4)
(4.5,4e-5)
};
\addlegendentry{ mRRD-RNN(5) \cite{nachmani18}} 
\addplot[
color=red,
mark=*,
solid,
very thin
]
coordinates {
(1.0, 0.0528013)
(1.5, 0.0265283)
(2.0, 0.0138031)
(2.2, 0.0104121)
(2.4, 0.0069783)
(2.6, 0.004969)
(2.8, 0.0028645)
(3.0, 0.0018165)
(3.2, 0.0010751)
(3.4, 0.0006603)
(3.6, 0.0003682)
(3.8, 0.0001911)
(4.0, 9.23e-05)
};	
\addlegendentry{NMS(4)+OSD(1)}
\addplot[
color=blue,
very thin
]
coordinates {
(1.0, 0.0450923)
(1.5, 0.0225135)
(2.0, 0.0097833)
(2.5, 0.0033765)
(3.0, 0.000807)
(3.5, 0.0002256)
};	
\addlegendentry{OSD(2)}
\addplot[
color=magenta,
mark=asterisk,
solid,
very thin
]
coordinates {
(3.0, 9e-4)
(3.5, 2.1e-4)
(4.0, 3e-5)
(4.5,3.9e-6)
};	
\addlegendentry{ML\cite{nachmani18}}
    \end{semilogyaxis}
\end{tikzpicture}
    }}
    \hspace{-0.025\linewidth}  
    \subfloat[BER of (63,45) code\label{fig:ber63_45}]{
    \resizebox{0.48\linewidth}{!}{    
 	\begin{tikzpicture}[scale=0.55]
		\begin{semilogyaxis}[
			scale = 0.75,
			xlabel={$E_b/N_0$(dB)},
			ylabel={BER},
			xmin=0.5, xmax=4.65,
			ymin=3e-5, ymax=0.12,
			xtick={0.5,1.0,1.5,2,2.5,...,4.0,4.5},
			legend pos = south west,
   		legend style={legend columns=1,fill opacity=0.6},
            draw opacity=1,text opacity=1,
			ymajorgrids=true,
            yminorgrids=true,
			xmajorgrids=true,
			grid style=dashed,
			]
\addplot[
color=black,
mark=diamond,
very thin
]
coordinates {
    (0.0,0.11)
    (0.5,0.098)
    (1.0,0.088)
    (1.5,0.07)
    (2.0,0.055)
    (2.5,0.04)
    (3.0,0.028)
    (3.5,0.018)
    (4.0,0.0095)
    (4.5,0.0048)
};
\addlegendentry{EPCM \cite{baldi2008iterative}}

\addplot[
color=teal,
mark=square,
very thin
]
coordinates {
(1.0,8e-2)
(2.0,5e-2)
(3.0,2.4e-2)
(4.0,7e-3)
(5.0,1.4e-3)
};
\addlegendentry{ BP-RNN 
\cite{nachmani18}}

\addplot[
color=cyan,
mark=triangle,
very thin
]
coordinates {
(3.0, 1.1e-2)
(3.25,6.8e-3)
(3.5,4.3e-3)
(3.75,2.5e-3)
(4.0,1.4e-3)
(4.25,7.5e-4)
(4.5,4.1e-4)
(5.0,1e-4)
};
\addlegendentry{mRRD(1)\cite{dimnik2009improved}}
\addplot[
color=violet,
mark=halfcircle,
solid,
very thin
]
coordinates {
(1.0, 0.07601)
(1.5, 0.05458)
(2.0, 0.03607)
(2.2, 0.0286)
(2.4, 0.02244)
(2.6, 0.01693)
(2.8, 0.01308)
(3.0, 0.00965)
(3.2, 0.00648)
(3.4, 0.00447) 
(3.6, 0.0028)
(3.8, 0.00186)
(4.0, 0.00116)
(4.2, 0.00071)
(4.4, 0.00039)
(4.6, 0.00022)
};	
\addlegendentry{NMS(4)}

\addplot[
color=red,
mark=triangle*,
very thin
]
coordinates {
(3.0, 6e-3)
(3.25,3.5e-3)
(3.5,2.05e-3)
(3.75,1.1e-3)
(4.0,5.3e-4)
(4.25,2.6e-4)
(4.5,1.2e-4)
(5.0,2.3e-5)
};	
\addlegendentry{mRRD(3) \cite{dimnik2009improved}}

\addplot[
color=magenta,
mark= +,
dashed,
thin
]
coordinates {
(0.0, 0.03206)
(0.5, 0.0264)
(1.0, 0.02137)
(1.5, 0.01708)
(2.0, 0.01108)
(2.2, 0.00885)
(2.4, 0.00699)
(2.6, 0.00539)
(2.8, 0.00386)
(3.0, 0.00303)
(3.2, 0.00215)
(3.4, 0.00153)
(3.6, 0.000936)
(3.8, 0.000577)
(4.0, 0.000405)
(4.2,0.000236)
(4.4,0.000138)
(4.6,0.000084)
};	
\addlegendentry{Undetected BER}

\addplot[
color=orange,
mark=x,
very thin
]
coordinates {
(3.0, 5e-3)
(3.25,2.7e-3)
(3.5,1.5e-3)
(3.75,7.4e-4)
(4.0,4e-4)
(4.25,1.8e-4)
(4.5,8e-5)
(5.0,1.3e-5)
};	
\addlegendentry{MBBP\cite{hehn2010multiple}}

\addplot[
color=purple,
mark= diamond*,
solid,
very thin
]
coordinates {
(3.0,3e-3)
(3.25,1.8e-3)
(3.5,1e-3)
(3.75,5.3e-4)
(4.0,2.9e-4)
(4.25,1.4e-4)
(4.5,6e-5)
};	
\addlegendentry{PBP\cite{ismail2015efficient}}
\addplot[
color=blue,
very thin
]
coordinates {
(1.0, 0.0614476)
(1.5, 0.040532)
(2.0, 0.0209087)
(2.5, 0.009601)
(3.0, 0.0037413)
(3.5, 0.0011048)
};	
\addlegendentry{OSD(2)}
    \end{semilogyaxis}
\end{tikzpicture}}
    }
    \caption{BER of various decoders for BCH codes of length 63} \label{fig:ber_63_36_45}
\end{figure}

\begin{figure}[!t]
    \centering
    \subfloat[BER of (127,64) and (127,78) codes \label{fig:ber127_64_78}]{
    \resizebox{0.49\linewidth}{!}{ 
	\begin{tikzpicture}[scale=0.55]
		\begin{semilogyaxis}[
			scale = 0.75,
			xlabel={$E_b/N_0$(dB)},
			ylabel={BER},
			xmin=1.0, xmax=6.1,
			ymin=1e-4, ymax=2e-1,
			xtick={1.0,...,4.0,5.0},
			legend pos = south west,
			ymajorgrids=true,
			xmajorgrids=true,
			grid style=dashed,
			legend style={legend columns=1},
            xminorgrids=false, 
            yminorgrids=true,
			]
\addplot[
color=orange,
mark=triangle,
thin
]
coordinates {
(1.0,0.14)
(2.0,0.11)
(3.0,0.071)
(4.0,0.04)
(5.0,0.02)
(6.0,0.0064)
};	
\addlegendentry{BP-64
\cite{nachmani18}}
\addplot[
color=purple,
mark=square,
thin
]
coordinates {
(1.0,0.14)
(2.0,0.11)
(3.0,0.071)
(4.0,0.04)
(5.0,0.013)
(6.0,0.003)
};
\addlegendentry{BP-RNN-64 \cite{nachmani18}}
\addplot[
color=magenta,
mark=halfcircle,
thin
]
coordinates {
(2.0, 0.1) 
(2.5, 0.085) 
(3.0, 0.07) 
(3.5, 0.05)
(4.0, 3e-2)
(5,1e-2)
};
\addlegendentry{RNN-SS+PAN-64 \cite{lian2019learned}}

\addplot[
color=red,
mark=diamond,
thick
]
coordinates {
(1.0, 0.12724)
(1.5, 0.10879)
(2.0, 0.08352)
(2.5, 0.06151)
(3.0, 0.03689)
(3.5, 0.0181)
(4.0, 0.00596)
(4.2, 0.00371)
(4.4, 0.00204)
(4.6, 0.00108)
(4.8, 0.000541)
(5.0, 0.00024)
};
\addlegendentry{NMS(8)-64}

\addplot[
color=cyan,
mark=triangle*,
thin
]
coordinates {
(1.0,0.14)
(2.0,0.11)
(3.0,0.061)
(4.0,0.03)
(4.4,1.8e-2)
(5.0,7e-3)
(6.0,7e-4)
};
\addlegendentry{BP-RPCM-71\cite{baldi2008iterative}}

\addplot[
color=blue,
mark=diamond*,
thick
]
coordinates {
(1.0,0.10415)
(1.5, 0.08805)
(2.0, 0.06769)
(2.5, 0.0463)
(3.0, 0.02608)
(3.5, 0.01235)
(4.0, 0.00417)
(4.2, 0.002377)
(4.4, 0.00136)
(4.6, 0.00068)
};
\addlegendentry{NMS(8)-78}

\end{semilogyaxis}
\end{tikzpicture}
    }}
    \hspace{-0.05\linewidth}  
    \subfloat[FER of (127,99) code  \label{fig:fer_127_99}]{
    \resizebox{0.49\linewidth}{!}{  	\begin{tikzpicture}[scale=0.55]
		\begin{semilogyaxis}[
			scale = 0.75,
			xlabel={$E_b/N_0$(dB)},
			ylabel={FER},
			xmin=1.0, xmax=7.1,
			ymin=1e-6, ymax=1.,
			xtick={1.0,...,4.0,5.0,6.0,7.0},
			legend pos = south west,
			ymajorgrids=true,
			xmajorgrids=true,
			grid style=dashed,
			legend style={legend columns=2, fill opacity=0.6},
            draw opacity=1,text opacity=1,
                xminorgrids=false, 
                yminorgrids=true,
			]
\addplot[
color=red,
mark=diamond,
very thin
]
coordinates {
(2.5, 0.708667)
(3.0, 0.514)
(3.5, 0.304545)
(4.0, 0.131429) 
(4.5, 0.044336)
(5.0,0.01050667)
(5.5,0.00178267)
(6.0,0.00019100)
};
\addlegendentry{NMS(8)}

\addplot[
color=teal,
mark=+,
very thin
]
coordinates {
(1.80,  1.326e-01)
(2.00,  8.921e-02)
(2.20,  6.061e-02)
(2.40,	4.340e-02)
(2.60,	2.864e-02)
(2.80,	1.674e-02)
(3.00,	1.074e-02)
(3.20,	6.122e-03)
(3.40,	3.996e-03)
(3.60,	2.670e-03)
(3.80,	1.914e-03)
(4.00,	1.261e-03)
(4.20,	8.035e-04)
(4.40,	5.684e-04)
(4.60,	3.793e-04)
(4.80,	2.097e-04)
(5.00,	1.235e-04)
(5.20,	6.470e-05)
(5.40,	3.589e-05)
(5.60,	2.350e-05)
(5.80,	1.207e-05)
(6.00,	6.530e-06)
(6.20,	3.932e-06)
(6.40,	2.359e-06)
(6.60,	1.348e-06)
(6.80,	6.880e-07)
(7.00,	2.737e-07)
};
\addlegendentry{ML-5G(128,96) \cite{helmling19}}

\addplot[
color=magenta,
mark=square,
thick
]
coordinates {
(2.5, 0.1087842)
(3.0, 0.0318484)
(3.5, 0.0063099)
(4.0, 0.0008574)
(4.5, 7.6e-05)
};
\addlegendentry{NMS(8)+OSD(2)}

\addplot[
color=violet,
mark=halfcircle,
very thin
]
coordinates {
(2.00,	3e-01)
(2.50,	1.e-01)
(3.00,	3e-02)
(3.50,	6e-03)
(4.0,   8e-4)
(4.5,   7.5e-5)
};	
\addlegendentry{HOSD\cite{bailon2022concatenated}}
 
\addplot[
color=blue,
mark=*,
very thin
]
coordinates {
(1.0, 0.48424)
(1.5, 0.33734)
(2.0, 0.19333)
(2.5, 0.084)
(3.0, 0.03017)
(3.5, 0.00637)
(4.0, 0.00086)
(4.5,8.2e-05)
};	
\addlegendentry{OSD(2)}
\addplot[
color=magenta,
mark=+,
very thin,
]
coordinates {
(2.00,	2.551e-01)
(2.50,	1.020e-01)
(3.00,	2.680e-02)
(3.50,	6.907e-03)
(4.00,	8.074e-04)
(4.50,	4.958e-05)
(5.00,	2.984e-06)
};
\addlegendentry{ML\cite{helmling19}}
\end{semilogyaxis}
\end{tikzpicture}}
    }  
    \caption{FER for length-127 codes with varying rates}
\label{fig:fer_ber_127}
\end{figure}

For the BCH (63,36) code at a BER of $10^{-3}$, Fig.~\ref{fig:ber63_36} shows that the NMS slightly outperforms the RNN-SS \cite{lian2019learned} and leads the GPT \cite{babalola2019generalized} and BP-RNN decoders \cite{nachmani18} by at least 1.5 dB. The 0.4 dB gap to the mRRD-RNN(5) \cite{nachmani18} can be bridged by employing the hybrid of NMS and order-1 OSD. For the BCH (63,45) code in Fig.~\ref{fig:ber63_45}, both the EPCM \cite{baldi2008iterative}, which exploits circulant PCM, and the BP-RNN are inferior to the NMS, highlighting the importance of PCM optimization. The NMS slightly outperforms the mRRD(1) but lags behind high-complexity ensemble decoders such as mRRD(3) and MBBP \cite{hehn2010multiple}, as well as the PBP \cite{ismail2015efficient}, a serialized variant of the mRRD(q) decoder. For both codes, the order-2 OSD can be considered as the ML upper bound, indicating that the hybrid NMS(4)+OSD(2) decoder achieves near-ML performance. Notably, for the high-rate (63,45) short code with a small minimum distance, an undetected BER curve limits the NMS's improvement due to false positive decodings.

For longer BCH codes of length 127, as shown in Fig.~\ref{fig:ber127_64_78}, the NMS significantly outperforms all other parallelizable decoders. At a BER of $10^{-3}$, the NMS leads the standard BP, BP-RNN, and RNN-SS+PAN \cite{lian2019learned} by at least 1.5 dB for the (127,64) code. Similarly, the NMS outperforms the BP-RPCM \cite[Fig.5]{baldi2008iterative}, designed for the lower-rate (127,71) code, by approximately 1.3 dB for the (127,78) code. Regarding FER comparison for the (127,99) code in Fig.~\ref{fig:fer_127_99}, the proposed hybrid of NMS and order-2 OSD overlaps with the HOSD \cite{bailon2022concatenated}, order-2 OSD alone, and its simulated ML curve. However, it stands out uniquely in terms of throughput and latency analysis, as discussed below. Interestingly, the NMS alone for the BCH (127,99) code is expected to surpass the ML performance of the lower-rate 5G LDPC (128,96) code \cite{helmling19} in the SNR $> 6.5$ dB region, as evidenced by the steeper slope of the former in Fig.~\ref{fig:fer_127_99}. Thus, benefiting from its superior minimum distance, BCH codes exhibit better FER potential than LDPC codes in high-SNR scenarios. Further simulations show that the hybrid of NMS and order-1 OSD achieves near-ML performance for BCH codes of length 63 with rates above half, while order-2 OSD is required for BCH codes of length 127 to achieve near-ML performance.

\subsection{Complexity Analysis}

\begin{table}[!t]
\caption{\scriptsize{\uppercase{Settings and Complexity for Decoders of (63,45) Code.}}}
\label{tab:complexity-table}
\resizebox{0.49\textwidth}{!}{%
\begin{tabular}{|c|ccc|c|}
\hline
\multirow{2}{*}{Decoders} &
  \multicolumn{3}{c|}{Settings} &
  \multirow{2}{*}{\begin{tabular}[c]{@{}c@{}}Complexity\\ Ratios\end{tabular}} \\ \cline{2-4}
 &
  \multicolumn{1}{c|}{($I_1,I_2,I_3$)} &
  \multicolumn{1}{c|}{\begin{tabular}[c]{@{}c@{}}\# of automorphisms\\ per iteration\end{tabular}} &
  $M_s$ of  PCM&
   \\ \hline
mRRD(q)\cite{dimnik2009improved} & \multicolumn{1}{c|}{(15,50,q)}       & \multicolumn{1}{c|}{varied one}           & 18 & q      \\ \hline
PBP\cite{ismail2015efficient}     & \multicolumn{1}{c|}{(15,50,q)}       & \multicolumn{1}{c|}{varied one}           & 18 & q      \\ \hline
MBBP\cite{hehn2010multiple}    & \multicolumn{1}{c|}{(66,1,3)}        & \multicolumn{1}{c|}{fixed one}            & 63 & 0.92   \\ \hline
BP-RNN\cite{nachmani18}  & \multicolumn{1}{c|}{$I_m=5$}           & \multicolumn{1}{c|}{fixed one}            & 18 & 0.0067 \\ \hline
EPCM\cite{baldi2008iterative}    & \multicolumn{1}{c|}{$I_m=5$} & \multicolumn{1}{c|}{fixed one}            & 63 & 0.023   \\ \hline
\color{blue}Enhanced NMS  & \multicolumn{1}{c|}{\color{blue}$I_m=4$}           &\multicolumn{1}{c|}{\color{blue}$|S_n||S_p|$= 9} & \color{blue}33 & \color{blue}0.088  \\ \hline
\end{tabular}
}
\end{table}

The computational complexity and key parameter settings for decoders applied to the BCH (63,45) code are presented in Table~\ref{tab:complexity-table}. The first three decoders require a total of $I_1I_2I_3$ BP iterations per sequence in the worst-case scenario. Specifically, the mRRD(q) with $q$ subdecoders in parallel necessitates $750q$ iterations. In comparison, the PBP operates similarly to the mRRD(q) but in a serial mode. The MBBP requires 198 iterations but suffers from high hardware and computational complexity due to the squared shape of its three deployed PCMs. Although the BP-RNN and EPCM schemes require only 5 BP iterations, their FER performance is far from competitive, as shown in Fig.~\ref{fig:ber63_45}. The NMS requires 9 automorphisms per iteration on a slightly redundant PCM, compared to one random or identity automorphism for other schemes. Ignoring the complexity discrepancy between BP and NMS iterations, the computational complexity of an iterative decoder is roughly proportional to the product of $|S_n||S_p| \ast I_m \ast M_s$. The complexity ratio, defined as the ratio of the current decoder's complexity to that of the benchmarked mRRD(1), is shown in the last column of Table~\ref{tab:complexity-table}. Combined with the FER results in Fig.~\ref{fig:ber63_45}, it is evident that the proposed NMS offers a better trade-off between FER performance and computational complexity than other decoders.

For a hybrid decoder, such as the proposed one or the HOSD \cite{bailon2022concatenated}, where the second component is triggered only if the first fails, the comprehensive FER is the product of the FERs of the two components. However, the total complexity $C_t = C_1 + F_1 \ast C_2$, where $C_2$ is weighted by the FER of the first component $F_1$. Notably, $C_t$ can broadly refer to time latency, throughput, etc. Evidently, $C_t$ approaches $C_1$ when $F_1$ is sufficiently small, underscoring the need for a competitive first-component decoder, which justifies the design of the proposed NMS. Compared to the algebraic decoder (the first component of the HOSD), the enhanced NMS dominates the decoding task over a much wider SNR range, resulting in fewer invocations of the serial-processing OSD. Therefore, due to the parallelizability and low complexity of the NMS, it can be inferred that the proposed hybrid outperforms the HOSD in terms of throughput, despite their comparable FERs.
\section{Conclusion}
\label{conclusion}

To the best of the authors' knowledge, this work presents the first parallelizable and effective decoder for short BCH codes. First, the adaptation of the parity-check matrix is facilitated by a heuristic method, achieving targets such as appropriate redundancy and reduction of length-4 cycles. Second, an enhanced NMS decoder is designed to leverage the cyclic properties of BCH codes, significantly enhancing both decoding performance and convergence speed. Additionally, we have identified potential risks to the FER performance due to rank deficiency inherent in the code structure, as well as false positive decodings in very short BCH codes. Finally, the hybrid decoder combining the enhanced NMS with OSD demonstrates competitive FER performance compared to other decoders while excelling in throughput and latency metrics across the SNR region of interest.
With advancements in hardware, particularly graphics processing units, the parallelizable nature of the NMS makes its combination with OSD highly compelling for achieving URLLC.

\begin{thebibliography}{10}
\providecommand{\url}[1]{#1}
\csname url@samestyle\endcsname
\providecommand{\newblock}{\relax}
\providecommand{\bibinfo}[2]{#2}
\providecommand{\BIBentrySTDinterwordspacing}{\spaceskip=0pt\relax}
\providecommand{\BIBentryALTinterwordstretchfactor}{4}
\providecommand{\BIBentryALTinterwordspacing}{\spaceskip=\fontdimen2\font plus
\BIBentryALTinterwordstretchfactor\fontdimen3\font minus \fontdimen4\font\relax}
\providecommand{\BIBforeignlanguage}[2]{{%
\expandafter\ifx\csname l@#1\endcsname\relax
\typeout{** WARNING: IEEEtran.bst: No hyphenation pattern has been}%
\typeout{** loaded for the language `#1'. Using the pattern for}%
\typeout{** the default language instead.}%
\else
\language=\csname l@#1\endcsname
\fi
#2}}
\providecommand{\BIBdecl}{\relax}
\BIBdecl

\bibitem{gallager62}
R.~Gallager, ``Low-density parity-check codes,'' \emph{IRE Trans. Inf. Theory}, vol.~8, no.~1, pp. 21--28, 1962.

\bibitem{mackay96}
D.~J. MacKay and R.~M. Neal, ``Near shannon limit performance of low density parity check codes,'' \emph{Electron. Lett.}, vol.~32, no.~18, p. 1645, 1996.

\bibitem{jiang2004iterative}
J.~Jiang and K.~R. Narayanan, ``Iterative soft decoding of {R}eed-{S}olomon codes,'' \emph{IEEE Commun. Lett.}, vol.~8, no.~4, pp. 244--246, 2004.

\bibitem{jiang2006iterative}
------, ``Iterative soft-input soft-output decoding of {R}eed--{S}olomon codes by adapting the parity-check matrix,'' \emph{IEEE Trans. Inf. Theory}, vol.~52, no.~8, pp. 3746--3756, 2006.

\bibitem{halford2006random}
T.~R. Halford and K.~M. Chugg, ``Random redundant soft-in soft-out decoding of linear block codes,'' in \emph{Int. Symp. Inf. Theory}.\hskip 1em plus 0.5em minus 0.4em\relax IEEE, 2006, pp. 2230--2234.

\bibitem{ismail2015efficient}
M.~Ismail, S.~Denic, and J.~Coon, ``Efficient decoding of short length linear cyclic codes,'' \emph{IEEE Commun. Lett.}, vol.~19, no.~4, pp. 505--508, 2015.

\bibitem{santi2018decoding}
E.~Santi, C.~Hager, and H.~D. Pfister, ``Decoding {R}eed-{M}uller codes using minimum-weight parity checks,'' in \emph{Int. Symp. Inf. Theory (ISIT)}.\hskip 1em plus 0.5em minus 0.4em\relax IEEE, 2018, pp. 1296--1300.

\bibitem{babalola2019generalized}
O.~P. Babalola, O.~Ogundile, and D.~J. Versfeld, ``A generalized parity-check transformation for iterative soft-decision decoding of binary cyclic codes,'' \emph{IEEE Commun. Lett.}, vol.~24, no.~2, pp. 316--320, 2019.

\bibitem{dimnik2009improved}
I.~Dimnik and Y.~Be'ery, ``Improved random redundant iterative hdpc decoding,'' \emph{IEEE Trans. Commun.}, vol.~57, no.~7, pp. 1982--1985, 2009.

\bibitem{hehn2010multiple}
T.~Hehn, J.~B. Huber, O.~Milenkovic, and S.~Laendner, ``Multiple-bases belief-propagation decoding of high-density cyclic codes,'' \emph{IEEE Trans. Commun.}, vol.~58, no.~1, pp. 1--8, 2010.

\bibitem{geiselhart2021automorphism}
M.~Geiselhart, A.~Elkelesh, M.~Ebada, S.~Cammerer, and S.~ten Brink, ``Automorphism ensemble decoding of {R}eed--{M}uller codes,'' \emph{IEEE Trans. Commun.}, vol.~69, no.~10, pp. 6424--6438, 2021.

\bibitem{nachmani18}
E.~Nachmani, E.~Marciano, L.~Lugosch, W.~J. Gross, D.~Burshtein, and Y.~Be’ery, ``Deep learning methods for improved decoding of linear codes,'' \emph{IEEE J. Sel. Topics Signal Process.}, vol.~12, no.~1, pp. 119--131, 2018.

\bibitem{lian2019learned}
M.~Lian, F.~Carpi, C.~H{\"a}ger, and H.~D. Pfister, ``Learned belief-propagation decoding with simple scaling and {SNR} adaptation,'' in \emph{Int. Symp. Inf. Theory (ISIT)}.\hskip 1em plus 0.5em minus 0.4em\relax IEEE, 2019, pp. 161--165.

\bibitem{buchberger2020pruning}
A.~Buchberger, C.~H{\"a}ger, H.~D. Pfister, L.~Schmalen, and A.~G. Amat, ``Pruning neural belief propagation decoders,'' in \emph{Int. Symp. Inf. Theory (ISIT)}.\hskip 1em plus 0.5em minus 0.4em\relax IEEE, 2020, pp. 338--342.

\bibitem{lucas1998iterative}
R.~Lucas, M.~Bossert, and M.~Breitbach, ``On iterative soft-decision decoding of linear binary block codes and product codes,'' \emph{IEEE J. Sel. Areas Commun.}, vol.~16, no.~2, pp. 276--296, 1998.

\bibitem{yedidia2002generating}
J.~S. Yedidia, J.~Chen, and M.~P. Fossorier, ``Generating code representations suitable for belief propagation decoding,'' in \emph{Proc. Annu. Allerton Conf. Commun. Control and Comput.}, vol.~40, no.~1.\hskip 1em plus 0.5em minus 0.4em\relax IEEE, 2002, pp. 447--456.

\bibitem{kou01}
Y.~Kou, S.~Lin, and M.~P. Fossorier, ``Low-density parity-check codes based on finite geometries: a rediscovery and new results,'' \emph{IEEE Trans. Inf. Theory}, vol.~47, no.~7, pp. 2711--2736, 2001.

\bibitem{Fossorier1995}
M.~P. Fossorier and S.~Lin, ``Soft-decision decoding of linear block codes based on ordered statistics,'' \emph{IEEE Trans. Inf. Theory}, vol.~41, no.~5, pp. 1379--1396, 1995.

\bibitem{bossert2022hard}
M.~Bossert, R.~Schulz, and S.~Bitzer, ``On hard and soft decision decoding of {BCH} codes,'' \emph{IEEE Trans. Inf. Theory}, vol.~68, no.~11, pp. 7107--7124, 2022.

\bibitem{bailon2022concatenated}
D.~N. Bailon, M.~Bossert, J.-P. Thiers, and J.~Freudenberger, ``Concatenated codes based on the plotkin construction and their soft-input decoding,'' \emph{IEEE Trans. Commun.}, vol.~70, no.~5, pp. 2939--2950, 2022.

\bibitem{baldi2008iterative}
M.~Baldi, G.~Cancellieri, and F.~Chiaraluce, ``Iterative soft-decision decoding of binary cyclic codes,'' \emph{J. Commun. Softw. Syst.}, vol.~4, no.~2, pp. 142--149, 2008.

\bibitem{helmling19}
M.~Helmling, S.~Scholl, F.~Gensheimer, T.~Dietz, K.~Kraft, S.~Ruzika, and N.~Wehn, ``{D}atabase of {C}hannel {C}odes and {ML} {S}imulation {R}esults,'' \url{www.uni-kl.de/channel-codes}, 2019.

\end{thebibliography}

\newpage
\begin{center}
\textbf{IEEE Copyright Notice}
\end{center}

\begin{ttfamily}
© 2025 IEEE. Personal use of this material is permitted. Permission from IEEE must be obtained for all other uses, in any current or future media, including reprinting/republishing this material for advertising or promotional purposes, creating new collective works, for resale or redistribution to servers or lists, or reuse of any copyrighted component of this work in other works.
\end{ttfamily}

\begin{center}
\textbf{Published Version}
\end{center}

This article has been accepted in IEEE Communications Letters. The final published version will appear soon.
\end{document}